\documentclass[aps,eqsecnum,prb,showpacs]{revtex4}
\def\eq#1{Eq.~(\ref{eq:#1})}
\def\sec#1{Section \ref{sec:#1}}

\usepackage[dvips]{graphicx,color}
\usepackage{epsfig}

\begin{document}

\title{Analysis of the contributions of three-body potentials in the equation of state of $^4{\rm He}$.}

\author{Sebastian Ujevic\footnotemark[1] and S.~A.~Vitiello\footnotemark[2]}

\affiliation{Instituto de F\'\i sica, Universidade Estadual de Campinas, 13083 Campinas - SP, Brazil \footnotetext[1]{{\rm e-mail: }sut@ifi.unicamp.br} \footnotetext[2]{{\rm e-mail: }vitiello@ifi.unicamp.br}}

\begin{abstract}

The effect of three-body interatomic contributions in the equation of state of $^4{\rm He}$ are investigated.  A recent two-body potential together with the Cohen and Murrell (Chem. Phys. Lett. 260, 371 (1996)) three-body potential are applied to describe bulk helium. The triple-dipole dispersion and exchange energies are evaluated subjected only to statistical uncertainties. An extension of the diffusion Monte Carlo method is applied in order to compute very small energies differences. The results show how the three-body contributions affects the ground-state energy, the equilibrium, melting and freezing densities.

\end{abstract}


\maketitle

\section{Introduction} \label{sec:int}

The unique properties of the helium systems at low temperature have attracted a continuous experimental and theoretical interest in the investigation of their ground state potential energy.\cite{jan95} In the past, the construction of the best potentials used semi-empirical methods where some parameters were obtained by fits to experimental data.  One of them, the so called HFDHE2 potential of Aziz and collaborators,\cite{azi79} has allowed the understand of many properties of helium in the condensed phases.\cite{kal81,cep86} Despite of small inconsistencies in this potential, it was used during a long time in these studies.  It was attractive to use an effective pair-wise additive potential and avoid considering high-order interactions among the atoms.

In the last decade, after a bound helium dimer was observed,\cite{luo96,sch96} great efforts were applied to develop {\sl ab initio} methods in the description of He-He potentials.\cite{azi95,kor97} This approach was very successful; interacting energies calculated using infinite order symmetry adapted perturbation theory, Green's function Monte Carlo results and accurate dispersion coefficients were fitted to a Tang-Tonnies model \cite{tan84} and produced \cite{kor97} to date the best characterization of the helium potential energy.\cite{azi97}

The use of very accurate two-body potential energies, like those offered by the {\sl ab initio} potentials, in the investigation of bulk helium, uncovered what was known for a long time: the correct description of many of its properties requires more general many-body potential. Among the many recent works where this situation was observed, we cite Refs.[\onlinecite{vit02,mor00,vit99}].

The first attempt to incorporate long range three-body interactions were made by Axilrod-Teller\cite{axi43} and Muto.\cite{mut43} They used third order perturbation theory to calculate the triple-dipole dispersion energy for spherical symmetric atoms. Effects of three-body exchange of electrons in trimers of helium started 10 years latter by Rosen\cite{ros53} using a valance bond approach. Since the work of Jansen and collaborators \cite{jan62,jan67} it is conjectured that three-body exchange energy is needed to understand the energy difference between the fcc and hcp crystalline structure. Many developments occurred in the investigation of nonadditive effects are reviewed in Ref.[\onlinecite{elr94}].

It is known that two-body interactions favor the hcp over the fcc structure. The hcp crystalline lattice is about 0,01\% more stable than the corresponding fcc structure.\cite{nie76} The two-body interaction potentials of the rare gas in general are very well known and the residuals errors associated to them can not be responsible for discrepancies in the calculated ground state energy and in the favored crystalline structure. At low temperatures all rare gas solids, but helium, crystallizes in a fcc lattice. To overcome these discrepancies higher order interactions need to be considered. The inclusion of triple-dipole interaction improved the agreement between the experimental and theoretical values of the ground state energy, but still left the hcp as the more estable structure. The inclusion of higher order terms in the dispersion energy like dipole-octopole and quadrupole-quadrupole terms did not modify the theoretical results significantly. So far, the proposed higher-order interactions greater than three-body have given no significant contributions to the interatomic potential. In this context it is important to better understand the three-body exchange interaction and have at the same time reliable ways to compute its associate energy.

As already mentioned, for the helium systems the largest and most well know part of the three-body interatomic potential is the triple-dipole term of Axilrod-Teller-Muto. The exchange contribution is less known. It is much more weak than the dispersion energy and might be of importance in the crystallization process. Despite of its importance, still now, competing calculations might differ by an order of magnitude \cite{lot00} and different potentials forms are used for fitting theoretical results obtained in calculations of the exchange energy.

Quantum Monte Carlo methods, where quantities of interest can be computed subject only to statistical uncertainties, can be very useful in the analysis and understanding of the different contributions to the potential energy. However straightforward use of these methods to compute small energies differences might not be possible. Results of independent runs and their associate statistical uncertainties might render a simple energy difference meaningless. Perturbation methods used together with quantum Monte Carlo methods still needs extrapolation \cite{whi79} that introduces further uncertainties.

To be able to compute energies subject only to statistical uncertainties (avoiding extrapolation) is not only a matter of principle but a necessity in the present case, where we have a delicate balance between energies. A better understanding of the individual contributions of the three-body interactions are important by themself and moreover can increase the physical content of the analytical functional forms used to fit their contributions. 

In this work, we have applied a recent two-body potential \cite{azi95} together with the three-body potential of Cohen and Murrell \cite{coh96} to describe properties of bulk $^4{\rm He}$. By developing an extension of the diffusion Monte Carlo (DMC) method we are able to compute and analyze the individual contributions of the Coulomb and exchange terms of the three-body interactions.The effect of these contributions are considered in the equation of state and quantitative results show how they affect the ground-state energy, the equilibrium, melting and freezing densities.

The paper is organized as follows, in the next section we present the Hamiltonian together with the interacting potentials used in this work. In \sec{dif} we briefly describe one of the standard implementation of the Diffusion Monte Carlo algorithm, then we present an extension of this algorithm. It allows the use of a single set of walkers and reweighting to compute properties of a system of helium atoms described by different interacting potentials. Section \ref{sec:sim} contains details of our simulations; the results are presented in \sec{res}. A discussion in \sec{dis} concludes the work.

\section{The Model} \label{sec:mod}

The Hamiltonian we use to describe the system of helium atoms is given by

\begin{equation} \label{eq:esch}
H = -{\hbar^2 \over{2m}} \nabla^2_R + V(R), 
\end{equation}

\noindent where $R=\{{\bf r_1}, {\bf r_2}, \ldots, {\bf r_N}\}$ stands for the $N$ coordinates of the helium atoms and $V(R)$ is the interatomic potential. In this work three sets of calculations were performed. In the first one the interatomic potential employed is an additive pair-wise potential $V_2(R)$ as proposed by Aziz and co-works.\cite{azi95} In a second set, we considered the $V_{2D}$ potential

\begin{equation} \label{eq:v2d}
V_{2D}(R) = V_2(R) + V_D(R),
\end{equation}

\noindent obtained by adding to $V_2$ a damped Axilrod-Teller-Mutto \cite{axi43,mut43,coh96} triple-dipole term ($ddd \equiv D$)

\begin{equation} \label{eq:d}
V_D = Z^{(3)}(111) {1+3\cos(\gamma_1)\cos(\gamma_2) \cos(\gamma_3)\over{(r_{12}r_{13}r_{23})^3}} F(r_{12},r_{13},r_{23}),
\end{equation}

\noindent where $Z^{(3)}(111)$ is a constant, the $\gamma_i$ are the internal angles of the triangle (formed by the three particles) and the $r_{ij}$ the lengths of its sides. The damping $F$ is given by the product

\begin{equation}
F(r_{12},r_{13},r_{23})=f(r_{12})f(r_{13})f(r_{23})
\end{equation}

\noindent that depends on

\begin{equation} f(r_{ij}) =\left\{
\begin{array}{ll}
\exp(-l(\frac{k}{r_{ij}} - 1)^2) & \mbox{if r $<$ k}\\
    1 & \mbox{otherwise,}
\end{array}
\right.
\end{equation}

\noindent where $l$ and $k$ are parameters. The damping of the dispersion energies at distances where charge overlap is significant is need for a reasonable description of the short range forces. The value used for $Z^{(3)}(111)$ in \eq{d} is 0.324 K, as obtained by double perturbation theory.\cite{dut71}

Finally, the most complete interatomic potential we have considered,

\begin{equation} \label{eq:v2dj}
V_{2DJ}(R) \equiv V_2(R) + V_D(R) + V_J(R),
\end{equation}

\noindent includes contributions from the exchange potential $V_J$ as well. The two last terms of \eq{v2dj} form the three-body potential proposed by Cohen and Murrell.\cite{coh96} The potential $V_J$ is expressed through symmetry adapted coordinates $Q_i$ (linear combinations of the three distances $r_{ij}$)

\begin{eqnarray}
Q_1 & = & \frac{1}{\sqrt 3}(r_{12}+r_{13}+r_{23}), \nonumber \\
Q_2 & = & \frac{1}{\sqrt 2}(r_{13}-r_{23}), \nonumber \\
Q_3 & = & \frac{1}{\sqrt 6}(2r_{12}-r_{13}-r_{23}).
\end{eqnarray}

\noindent Its functional form is given by:

\begin{eqnarray} \nonumber
V_{J} &=& [c_0+c_1Q_1+c_2Q_1^2+ 
\\ && (c_3+c_4Q_1+c_5Q_1^2)(Q_2^2+Q_3^2)+ \nonumber
\\ && (c_6+c_7Q_1+c_8Q_1^2)(Q_3^3-3Q_3Q_2^2)+ \nonumber
\\ && (c_9+c_{10}Q_1+c_{11}Q_1^2)(Q_2^2+Q_3^2)^2+ \nonumber
\\ && (c_{12}+c_{13}Q_1+c_{14}Q_1^2)(Q_2^2+Q_3^2) \times \nonumber
\\ && (Q_3^3-3Q_3Q_2^2)]{\rm exp}(-\alpha Q_1),
\end{eqnarray}

\noindent that depends on parameters $\alpha$ and $\{c_i \mid i=1,\ldots,14\}$. 

\section{The Diffusion Monte Carlo Method} \label{sec:dif}

\subsection{The Standard Algorithm}

In almost all practical implementations of the diffusion Monte Carlo method \cite{mos82,rey82} we compute quantities of interest by sampling the probability distribution $f_0(R)=\psi_G(R)\psi_0(R)$ that depends on $\psi_0(R)$, the true ground state wave function of the system and on a given guiding function, $\psi_G(R)$.

A careful version of this method is presented by Umrigar, Nightingale and Runge.\cite{umr93} A slight simpler implementation could be described as follow. It is convenient to start the calculation with a set of configurations draw from $|\psi_G|^2$, obtained through the Metropolis algorithm.  The distribution $f_0(R)$ is sampled after an initial transient, where the excited states components are filtered. All the sampling is accomplished iteratively through the integral equation

\begin{equation} \label{eq:it}
f(R,\tau) = \int dR^{'} \, G_d(R,R^{'})G_b(R,R^{'}) f(R^{'}, \tau-\triangle \tau),
\end{equation}

\noindent where

\begin{equation} 
G_d(R,R^{'}) = (4 \pi D \triangle\tau)^{-\frac{3N}{2}} \exp \left[ -\frac{(R - R^{'}-D\triangle\tau v_D(R^{'}))^2}{4D\triangle\tau} \right],
\end{equation}

\begin{equation} 
v_D= 2{\bf \nabla} \ln \Psi_G,
\end{equation}

\begin{equation} \label{eq:gb}
G_b(R,R^{'}) = \exp\{- (\frac{\triangle\tau}{2}) [E_L(R) + E_L(R^{'})] + \triangle\tau E_T\} ,
\end{equation}

\noindent here $E_T$ is a trial energy and $E_L$ is the local energy given by

\begin{equation} \label{eq:le}
E_L={H \Psi_G}/{\Psi_G};
\end{equation}

\noindent $f(R,\tau)$ for a long enough $\tau$ goes to $f_0(R)$.

It is correct to write the Green's function as the product of \eq{it} only up to ${\cal O}(\tau^3)$. This implies that to obtain exact results, within statistical fluctuations, short times $\Delta \tau$ must be used in the iterations and an extrapolation to $\Delta \tau \rightarrow 0$ performed (see however comments in \sec{sim}). This is the so called short-time approximation.

Each configuration undergoes three steps: drift, diffusion and branching. Very frequently a single configuration is called a walker and an iteration of all walkers a generation. For the drift step we need to compute the quantum velocity $v_D$. In the second step the configuration diffuses. This is accomplished by sampling $G_d.$ Accordingly, a walker in $R'$ is propagated during a time step $\Delta \tau$ to its new point $R$ through

\begin{equation} \label{eq:nr}
R =  R' + \chi + D \Delta \tau  v_D( R'),
\end{equation}

\noindent where ${\bf \chi}$ are normal deviates of a Gaussian function with variance $2D \Delta \tau$ and zero mean.

To propagate a walker, we can change all the particle's coordinates at once or those of a single particle at a time. In this last case we perform N updates to propagate each walker. To improve \cite{cep81,rey82} the approximation of the Green's function, we only accept moves with probability

\begin{equation} \label{eq:db}
p_{\rm accept}(R^{'}\rightarrow R)={\rm min}\left[1,\frac{G_d(R^{'},R)\Psi_G^2(R)}{G_d(R,R^{'})\Psi_G^2(R^{'})}\right]
\end{equation}

\noindent and chose $\Delta \tau$ such that more than 99\% of the attempted moves are accepted. This condition imposes detailed balance on the splited Green's function and restores this property of the exact Green's function. Regardless of the time step, it guarantees also a correct sampling if hypothetically we could use as $\psi_G$ the exact ground state wavefunction. In such case, this implementation of the DMC method reduces to a variational Monte Carlo calculation with trial moves sampled from $G_d$.

To complete the iteration of \eq{it}, we compute $G_b$ considered as a weight for the walker. At the begin of the simulation, all the walker's weights are assumed to be equal to one. In order to minimize fluctuations in $G_b$, an effective time step is used.\cite{rey82,umr93} It is given by $\Delta\tau_{eff}=\Delta\tau (\Delta \rho^2_a/\Delta \rho^2)$, where $\Delta \rho^2$ is the mean square displacement of all proposed moves in the diffusion step and $\Delta \rho^2_a$ is the related quantity when only accepted moves are considered. Finally the weight $w'$ of the walker is updated to its new value $w$ according to

\begin{equation} \label{eq:nw}
w = w'G_b(R,R').
\end{equation}

After propagating all walkers we have a new generation and a sample of the probability distribution $f$ as an weighted average over the walkers. For reasons of efficiency the number of configurations used in the estimation of $f(R)$ fluctuates according to the following branching rules. If $w$ is greater than 2, the walker is duplicate and each one will carry half of its weight. On the other hand if two walkers, $R_i$ and $R_j$, have weights less than 0.5, only one survives with a weight given by $w_i+w_j$.  The decision of which one will survive is made by sampling $r=w_i/(w_i+w_j)$. That is, we draw a random number $\xi$ and compare it with $r$. If $r$ is small than $\xi$ we keep configuration $R_j$, otherwise $R_i$ is kept. If the weight computed by \eq{nw} lies between 0.5 and 2, a single copy of the walker with weight $w$ is kepted. The values of the weights where the branching is performed certainly could be changed, we only need rules that neither introduce bias or result in a scheme too inefficient.

The number of walkers is controlled by adjusting the value of $E_T$. This number is kept roughly constant. We have experienced with both heuristic and automatic changes of $E_T$ as given by

\begin{equation}
E_T = E_0 + \kappa \ln (tp/cp),
\end{equation}

\noindent where $\kappa$ is a parameter, $tp$ is the target population and $cp$ the current population. Adjustments in $E_T$ was made about once every 20 generations. For our purposes, the results were equivalent for both methods of changing $E_T$.

\subsection{The Algorithm with Reweighting} \label{sec:ce}

In many situations it is interesting to compute energies differences resulting from different interatomic potentials.  However it is not always possible to simply use results from independent runs to obtain such differences. If they are small, statistical fluctuations might well produce errors that are bigger than these differences themselves, rendering the result meaningless.  It is however possible to modify the DMC method in such a way that the same set of walkers are used to compute quantities of interest associated to the different potentials we want to investigate. The energies that are obtained are correlated and thus more meaningful differences can be computed. No approximations are introduced. If the actual interest is on the energies, it is not necessary to use extrapolated estimators either. What we are proposing is to sample different probabilities distributions functions, associated to the different interatomic potentials we want to investigate, by using the same set of walkers with appropriate weights. As just mentioned, the values of the quantities of interest obtained are correlated and the errors associate with their difference reduced by orders of magnitude. We want to call attention to the fact that although our method relay in a set of weights it can not be related to the forward walking \cite{liu74} or reptation \cite{bar99} methods. Always that we have a generation of equilibrated walkers, we can compute the quantities of interest without any further propagation. Moreover the weights we have for a given walker are associate to different interatomic potentials.

In our modified DMC method, to each walker we attach a set of weights, one for each potential we want to consider.  In our implementation of the algorithm we have attached three different weights for each walker, one for each of the three different interatomic potentials used. Of course this number in the method is arbitrary. It was chosen because of the specific aspects of many-body interactions in the interatomic potential we want to investigate. It would be possible to use only two weights or any other convenient number of weights.

As in the standard algorithm, the calculations start with a set of walkers draw from $|\psi_G|^2$. Since a single guide function is use, the drift and diffusion steps are performed exactly as before. We compute the drift velocity $v_D$, generate the normal distribution of variates, update $R$ according to \eq{nr} and accept it with probability $p_{accept}$ of \eq{db}. In the present algorithm, we sample the three different probabilities distributions in which we are interested by completing the iteration considering three different $G_b$, and updating the weights as follows.

To be specific let us consider a single walker just propagated to a new configuration $R$. One of its weights, $w^{(2)}$ is associate with the local energies $E^{(2)}_L(R)$ computed using only the Aziz two-body potential $V_2$ in the Hamiltonian of \eq{esch}. The weight $w^{(2)}$ is updated according to \eq{nw} by evaluating $G_b$ of \eq{gb} using the local energy $E^{(2)}_L(R)$. Another weight of the same walker, $w^{(2D)}$ is calculated with the local energy $E_L^{(2D)}(R)$ computed with the Hamiltonian that uses the $V_{2D}$ potential of \eq{v2d}, it includes the triple-dipole contributions to the two-body potential. The calculation of the new value of $w^{(2D)}$ for this walker proceeds as before. $G_b$ in \eq{gb} is evaluated employing $E_L^{(2D)}(R)$ and the update finished according to \eq{nw}. The third weight, $w^{(2DJ)}$, is associate to the interatomic potential that includes also exchange contributions. It is computed by considering $E_L^{(2DJ)}(R)$ that depends on the Hamiltonian that uses the full potential $V_{2DJ}$ of \eq{v2dj}. The update of $w^{(2DJ)}$ is performed along the exact same lines already describe for the other weights. The different values of the weights are due only to the local energy used in their computation, {\sl i. e.}, to the interatomic potential employed. We remember again that the same configuration is used to compute these three weights.

After propagating all walkers, a new generation will finally be obtained by using the following branching rules. When $\min(w^{(2)},w^{(2D)},w^{(2DJ)})$ is large than 2 the walker is duplicated and each one of the copies will carry half of the value of the weights: $(w^{(2)}/2,w^{(2D)}/2,w^{(2DJ)}/2)$. If two walkers, $i$ and $j$ have weights such that $\max(w_i^{(2)},w_i^{(2D)},w_i^{(2DJ)})$ and $\max(w_j^{(2)},w_j^{(2D)},w_j^{(2DJ)})$ are less than 0.3, we consider each one of the weights individually. We draw a single random number $\xi$ and make comparisons of $\xi$ with $r^{(2)}$, $r^{(2D)}$ and $r^{(2DJ)}$, where $r^{(k)} = w^{(k)}_i/(w^{(k)}_i+w^{(k)}_j)$. Three situations might happen: i) in all the comparisons $\xi$ is smaller than $r^{(k)}$, then we keep walker $i$ with weights $\{w^{(2)}_i+w^{(2)}_j,\ w^{(2D)}_i+w^{(2D)}_j,\ w^{(2DJ)}_i+w^{(2DJ)}_j\}$ and discard walker $j$; ii) always $\xi$ is greater than $r^{(k)}$, in this case we keep $R_j$ with the same sum of weights as above and discard $R_i$; iii) one of the comparisons favors a walker different from the other two. For definiteness let say that $\xi$ is smaller than $r^{(2)}$ and greater than $r^{(2D)}$ and $r^{(2DJ)}$. In this case we will keep the two walkers, $R_i$ with weights $\{w^{(2)}_i+w^{(2)}_j,\ 0,\ 0\}$ and $R_j$ with weights $\{0,\ w^{(2D)}_i+w^{(2D)}_j,\ w^{(2DJ)}_i+w^{(2DJ)}_j\}$. These new weights are telling us that in fact we have deleted walker $i$ from the calculations with the interatomic potential that includes three-body interactions and walker $j$ when we are considering only the two-body potential. This is a bad situation in the sense that we are introducing two walkers in the calculations that will not give anymore the correlations that we are looking for. Fortunately, if needed, the cases where this situation happens can be systematically reduced in a simple way. It is enough to decrease the threshold value used to combine walkers (see \sec{sim}). Finally if one of the weights of a walker lies between 0.3 and 2, a single copy is maintained with weights $(w^{(2)},w^{(2D)},w^{(2DJ)})$.

It is useful to use only a single random number in the above comparisons. As expected each one of the calculations we are performing give exactly, within statistical fluctuations, the results obtained by the standard algorithm. Of course we could use three different random numbers in the comparison, however more frequently we would meet the unwanted situation described in iii) of the last paragraph.

Periodically, about one every four or five generations we compute several quantities of interest. Evaluations of the energies $E^{(2)}_m$, $E^{(2D)}_m$ and $E^{(2DJ)}_m$ are readily obtained as weighted averages that include all the walkers $R_i$ of the present generation:

\begin{eqnarray} \label{eq:me}
E^{(k)}_m = {\sum_i w^{(k)}(R_i)E_L^{(k)}(R_i) \over {\sum_i w^{(k)}(R_i)}}, \,\,\, (k=2,2D,2DJ).
\end{eqnarray}

\noindent Together with these quantities we have also evaluated the energy associate only to the damped triple-dipole term in the interatomic potential

\begin{equation} \label{eq:ed}
E^D_m = E^{(2D)}_m - E^{(2)}_m,
\end{equation}

\noindent and the energy associate with the exchange term $V_J$

\begin{equation} \label{eq:ej}
E^J_m = E^{(2DJ)}_m - E^{(2D)}_m.
\end{equation}

\noindent As already mentioned the computation of these values are straightforward because we have already estimates of the exact energies $E^{(k)}_m$, no extrapolations are needed. Along the runs, averages of these quantities are formed and their estimates and associate errors obtained.
 
\section{The Simulations} \label{sec:sim}

In the investigation of the properties of bulk helium we impose periodic boundary conditions. The cutoff convention, the distance beyond which a potential is set to zero, is enforced for all interactions  at half of the box size, $L/2$. Distances between pairs of particles are computed by the minimum-image convention. When considering three-body interatomic interactions, the length of the third side of the triangles formed by the particles can not in general be computed using the minimum-image convention. A modification needs to be introduced so that the length of this side can be computed in a proper way and discarded if greater than $L/2$. To be specific let us consider particles $i$, $j$ and $k$. We compute distances $r_{ij}$ and $r_{ik}$ using the minimum-image convention. The difference in the $x$ coordinates of the associated particles are

\begin{eqnarray} \nonumber
x_{ij} &=& x_i - x_j - t_{ij}, \\
x_{ik} &=& x_i - x_k - t_{ik},
\end{eqnarray}

\noindent where the translation vector $t$ is defined as

\begin{eqnarray} \nonumber
t_{lm} &=& [(x_l - x_m)/L]L
\end{eqnarray}

\noindent and $[x]$ is the closest integer to $x$. If the difference $x_{jk}$ of the third side is computed as \cite{att92}

\begin{eqnarray}
x_{jk} &=& x_j - x_k + t_{ij} - t_{ik},
\end{eqnarray}

\noindent it is not hard to see that all possibilities in the simulation box are taken into account and the right value of the distance can be obtained. If this value is little than $L/2$ the calculation for this triangle proceeds. For the $y$ and $z$ coordinates a similar approach is used and then the three-body interaction is computed if all sides for this triangle are lower than $L/2$. The calculation continues until all triangles have been considered.

The diffusion Monte Carlo calculations started with an initial set of 400 configurations, previously draw from $|\psi_G|^2$ using the Metropolis algorithm. Before accumulating quantities of interest the excited states components of our ensemble of configurations are filtered by performing several iterations of \eq{it}. This ``equilibration" is typically of the order of 400 generations, and depends on the system density.

We study the liquid phase using a guiding function of the Jastrow form

\begin{equation} \label{eq:j}
\Psi_{J}(R)=\prod_{i<j} f(r_{ij}),
\end{equation}

\noindent where the factor $f(r_{ij}) = \exp(-u(r_{ij})/2)$ explicitly correlates pairs of particles through a pseudopotential of the McMillan form $u(r_{ij})=(b/r_{ij})^5$; $b$ is a parameter.

For the solid phase we have used a Nosanov-Jastrow 

\begin{equation} \label{eq:nj}
\Psi_{NJ}(R)=\Psi_{J}(R)\Phi(R)
\end{equation}

\noindent guiding function, where

\begin{equation}
\Phi(R)=\prod_{i} \exp \left[ \frac{C}{2}({\bf r}_i - {\bf l}_i)^2 \right]
\end{equation}

\noindent is a mean field term that localizes the particles around the given lattice sites ${\bf l}_i$.

All guiding functions were previously optimized by performing variational calculations. Although this is a convenient way of obtaining the values of the parameters, in principle they could be obtained without performating such calculations. It would be enough to chose parameters values that give the fastest filtering of the excited states of the initial configurations.

The time steps $\Delta\tau$ used in the calculations depends on the density. Their values vary within the range 0.001 to 0.002 (${\rm K}^{-1}$) in order to obtain more than $99 \%$ of acceptance of the attempted moves. We also observed that at this acceptance level, the extrapolation to $\Delta \tau \rightarrow 0$ of the energies values were in excellent agreement, within statistical fluctuations, to the actual values obtained in the calculation itself.

The quantities reported in this work were obtained by forming averages with about 500 estimates. Each estimate was performed after 4 generations. Blocking was used in order to avoid correlations in the calculations of the variances. The number of walkers during the simulations did not change by more than 10\%.

We have considered systems with 108 particles in the solid phase. In the liquid phase we have considered 64 particles. At $\rho_0=21.86$ $\rm{nm}^{-3}$ to estimate size effects we have also performed simulations with 108 particles. Tail corrections of the two-body potential energy were made by assuming a pair distribution function equal to one beyond half the size of the simulation cell and integrating the potential up to infinity. No tail 
corrections were performed for the high-order interactions. For the Axilrod-Teller interaction, the tail correction is less than 7\% of its value at $\rho_0=21.86$ $\rm{nm}^{-3}$ (see next section and Table II). This value is in rough agreement with a previous estimate of this quantity.\cite{mur71} For the exchange energy the relative tail correction is bigger than the one of the dispersion energy. However it should remain within the statistical uncertainty of our results (see Tables I and II).  

The situation where we have a walkers with one of its weights equal to zero destroys the correlation we want to construct. If we combine walkers when all their weights is less or equal 0.3, we noticed that the number of walkers with at least one of weights equal zero does not exceed $2\%$ of their total number. If needed this fraction can be further and systematically reduced by using a threshold smaller than 0.3 to combine walkers. As we have observed in our calculations, this is done at expense of a less efficient calculation. We have concluded that the threshold 0.3 for combination of walkers is perfectly reasonable for our purposes.

\section{Results} \label{sec:res}

\subsection{Liquid phase}

We conducted several independent runs at four different densities $\rho$ of liquid helium, 19.64 $\rm{nm}^{-3}$, at the experimental equilibrium density $\rho_0=21.86$ $\rm{nm}^{-3}$, 24.01 $\rm{nm}^{-3}$ and at 26.23 $\rm{nm}^{-3}$. In Table I are shown the total energies obtained using the two-body potential $V_2$, the $V_{2D}$ potential of Eq.(\ref{eq:v2d}), the $V_2$ potential plus the Axilrod-Teller term, and the $V_{2DJ}$ potential of Eq.(\ref{eq:v2dj}), the exchange term added to the $V_{2D}$ potential. 

It is important to note that since the energies associated with these potentials are calculated with a single set of walkers, they are correlated. We can believe that the results show their evolution as more elaborated interacting potentials are used, despite of the statistical uncertainties in the results.

In Table II we shown very accurate calculations of the Axilrod-Teller and exchange energies at these four densities. For comparison we show also extrapolated results of perturbative calculations performed using configurations generated with the $V_{2}$ potential. We have plot these results in Figures 1 and 2. The Axilrod-Teller energies calculated using reweighting are greater than the extrapolated perturbative results. Moreover they do not always agree within the statistical uncertainty. The triple-dipole interaction gives a positive contribution to the energy of the system and its value double when we go from the lowest to the highest density.  

The energies due to the exchange term in the $V_{2DJ}$ potential when calculated with respect to the total energies obtained with the $V_{2D}$ potential are on average about 0.0010 K smaller than the extrapolated perturbative results. In addition there is no agreement within statistical uncertainties between the energies calculated with the reweighting and perturbative methods at $\rho=19.64$ and $\rho=24.01$ ${\rm nm}^{-3}$. The energies calculated using reweighting are lower than the extrapolated perturbative results, contrary to what happens with the triple-dipole interaction. The exchange energy is also positive at all densities examined and increases with it. At the highest density it is approximately 50\% greater than in the lowest one. 

\subsection{Solid phase}

For the solid phase we have considered four densities: 29.34, 32.88, 33.54 and 35.27 $\rm{nm}^{-3}$. In Table I are shown the total energies, obtained with a systems of 108 particles in a fcc structure. Again, we have considered the potentials $V_2$, $V_{2D}$ and $V_{2DJ}$. Table II shown our very accurate results of the Axilrod-Teller and exchange contributions to the potential energy and also extrapolated perturbative results for comparison. The difference is about the same we have observed in the liquid phase. For the Axilrod-Teller term they do not agree within the statistical uncertainty at $\rho$ equal to 32.88 and 35.27 ${\rm nm}^{-3}$. In this phase as well, the Axilrod-Teller energies computed by reweighting are greater than the corresponding extrapolated perturbative results, and they remain positive. They also increase with the density. At the highest density (35.27 $\rm{nm}^{-3}$) it is 60\% greater than in the lowest one.

The contribution of the exchange term in the solid region is null or negative and differs significantly from the perturbative results. At the lowest density (29.34 $\rm{nm}^{-3}$) the result obtained by reweighting gives a null contribution while the extrapolated perturbative quantity is positive. In the other densities the exchange energy computed by reweighting continues to be lower than the corresponding extrapolated perturbative values that are negative. In the solid phase the relative variation of the exchange energy is greater than the corresponding quantity for the damped Axilrod-Teller energy.  

\subsection{Melting-Freezing Transition}

In order to follow the variations of the melting and freezing densities with the interacting potentials, we used a Maxwell (double-tangent) construction in analytical equations of state for the liquid and crystalline phases. The equations were determined by fits of our results to functions of the form

\begin{eqnarray} \label{eq:eqes}
E(\rho)=E_0+B(\frac{\rho-\rho_0}{\rho_0})+C(\frac{\rho-\rho_0}{\rho_0}).
\end{eqnarray}

\noindent This functional form has been extensively used in the literature, including to fit experimental equation of state.\cite{azi73,roa70} We have fitted equations of state using results from the three different potentials, $V_2$, $V_{2D}$ and $V_{2DJ}$. The fitted parameters, $E_0$, $B$, $C$ and $\rho_0$ in the liquid and solid phases are presented in Table III for these potentials. In Fig. 3 we display results for the equations of state for the three potentials.

The freezing and melting densities determined by the Maxwell double tangent construction are listed in Table IV. Looking at this table, we can follow the changes in the freezing and melting densities as more elaborated interacting potentials are used. The computed freezing densities differs, about 3\% from the experimental value. This difference is of about 4\% for the melting densities. The calculated freezing densities are below the experimental value, contrary to the computed melting densities that are above the experimental value.

\section{Discussion} \label{sec:dis}

In this work we are able to verify without any approximations how small changes in the interacting potential affects some of the properties of bulk helium. It was possible to analyze how the Axilrod-Teller and the exchange three-body contributions to the interatomic potential modify the equation of state of this system. This was accomplished by using a single set of walkers with reweighting in a DMC calculation. The quantities of interest associate with the different potentials were obtained in a correlated fashion and so despite of the statistical errors their difference are meaningful.

The total energies per atom presented in Table I show us that since our two-body potential is very accurate, high order terms in the description of the atomic interaction are needed. In the same token, as the contributions of the dispersion energy are much better known than those of the exchange energy, the results suggest that more efforts would be desirable in developing reliable ways of computing the energies associated to this last kind of interaction. 

The results of the Axilrod-Teller triple-dipole dispersion energy and of the three-body exchange energy as a function of the density reported in Table II, Figures 1 and 2, show qualitative agreement between the two methods we use in the calculations. However the usual approach of extrapolating perturbative calculations (made with configurations generated by the DMC method using a two-body potential) not always can be trusted in giving the right magnitude of this quantity. In some cases the results do not agree within statistical uncertainty with those obtained using reweight. This might be due to the simple functions used for extrapolation: functions of the Jastrow and Jastrow-Nosanov form, for the liquid and solid phases respectively. As expected, in both liquid and solid phases, the Axilrod-Teller term gives more important energy contribution to the total energy than the exchange term. Also it is interesting to note that at the lowest solid density we have considered (29.34 ${\rm nm}^{-3}$), the extrapolated perturbative result associated with the exchange term gives a positive energy contribution whereas the value obtained by reweighting is null.

Although the inclusion of the three-body interaction potentials used in this paper do not greatly modifies the melting and freezing densities, it is important to note that our calculations show that the inclusion of the triple-dipole and the exchange terms, as proposed by Cohen and Murell,\cite{coh96} are leading the melting and freezing densities to their right values. All the small differences we observe in these quantities are within their estimate error (0.2 $\rm{nm}^{-3}$). However since we used a single set of walkers to compute them, we can trust that their relative difference are significant in our calculations. Other consequence of including many body interactions in the potential can be seen in the calculation of the equilibrium density (Table III, \eq{eqes}). For both potentials, $V_{2D}$ and $V_{2DJ}$, the theoretical computed value of the equilibrium density becomes almost identical to the experimental value. The equilibrium density obtained using the full interatomic potential $V_{2DJ}$ diminished 0.3 ${\rm nm}^{-3}$ from its value computed with the two-body potential. In the solid phase, in which the contribution of the exchange energy is greater, the parameter $\rho_0$ decreases 0.75 ${\rm nm}^{-3}$ in a similar comparison.

The inclusion of three-body terms in the interatomic potential improves the agreement between computed and experimental values of properties like the binding energy, the equilibrium, melting and freezing densities. The results suggest that more reliable analytical expressions are needed for the calculation of the exchange energy in bulk helium. We could reach these conclusions by introducting reweighting in a DMC calculation, where a single set of walkers is used to compute properties associated to different potentials. Moreover we are able to calculate the energies associated to these potentials without extrapolation.

The extension we have proposed to the DMC method might be very useful not only for the helium systems, but also for other quantum many-body systems where a clue is need to identify the best description between competing interacting potentials. Even if these potentials differ by very small amounts, the nature of the interactions they describe can be different. A better understanding of these differences and their relevance will enlarge our knowledge about the interactions and so about these systems themselves. The questions discussed in this work are not the only instance where the reweighting technique might be useful. The calculations of small energy contributions of spin-orbit terms in molecular physics\cite{sar03} might be another situation where this method can help in a better understanding of a physical system. 

\acknowledgments

This work was conducted, in part, using the facilities of the ``Centro Nacional de Processamento de Alto Desempenho em S\~ao Paulo". SU thanks a fellowship from ``Funda\c{c}\~ao de Amparo \`a Pesquisa do Estado de S\~ao Paulo - FAPESP".

%\bibliography{nsav}

\begin{thebibliography}{99}

\bibitem{jan95} A.~R.~Janzen and R.~A.~Aziz, J. Chem. Phys. {\bf 103}, (1995) 9626.
\bibitem{azi79} R.~A.~Aziz, V.~P.~S.~Nain, J.~S.~Carley, W.~L.~Taylor and G.~T.~McConville, J. Chem. Phys. {\bf 70}, (1979) 4330.
\bibitem{kal81} M.~H.~Kalos, M.~A.~Lee, P.~A.~Whitlock and G.~V.~Chester, Phys. Rev. B {\bf 24}, (1981) 115.
\bibitem{cep86} D.~M.~Ceperley and E.~L.~Pollock, Phys. Rev. Lett. {\bf 56}, (1986) 351.
\bibitem{luo96} F.~Luo, C.~F.~Giese and W.~R.~Gentry, J. Chem. Phys. {\bf 104}, (1996) 1151.
\bibitem{sch96} W.~Schollkopf and J.~P.~Toennies, J. Chem. Phys. {\bf 104}, (1996) 1155.
\bibitem{azi95} R.~A.~Aziz, A.~R.~Janzen and M.~ R.~Moldover, Phys. Rev. Lett. {\bf 74}, (1995) 1586.
\bibitem{kor97} T.~Korona, H.~L.~Williams, R.~Bukowski, B.~Jeziorski and K.~Szalewicz, J. Chem. Phys. {\bf 106}, (1997) 5109.
\bibitem{tan84} K.~T.~Tang and J.~P.~Toennies, J. Chem. Phys. {\bf 80}, (1984) 3726.
\bibitem{azi97} A.~R.~Janzen and R.~A.~Aziz, J. Chem. Phys. {\bf 107}, (1997) 914.
\bibitem{vit02} S.~A.~Vitiello, Phys. Rev. B {\bf 65}, (2002) 214516.
\bibitem{mor00} S.~Moroni, F.~Pederiva, S.~Fantoni and M.~Boninsegni, Phys. Rev. Lett. {\bf 84}, (2000) 2650.
\bibitem{vit99} S.~A.~Vitiello and K.~E.~Schmidt, Phys. Rev. B {\bf 60}, (1999) 12342.
\bibitem{axi43} B.~M.~Axilrod and E.~Teller, J. Chem. Phys. {\bf 11}, (1943) 299.
\bibitem{mut43} Y.~Muto, Proc. Phys. Math. Soc. Jpn. {\bf 17}, (1943) 629.
\bibitem{ros53} P.~Rosen, J. Chem. Phys. {\bf 21}, (1953) 1007.
\bibitem{jan62} L.~Jansen, Phys. Rev. {\bf 125}, (1962) 1798.
\bibitem{jan67} L.~Jansen and E.~Lombardi, Chem. Phys. Lett {\bf 1}, (1967) 33.
\bibitem{elr94} M.~J.~Elrod and R.~J.~Daykally, Chem. Rev. {\bf 94}, (1994) 1975.
\bibitem{nie76} K.~F.~Niebel and J.~A.~Venables, in ``Rare Gas Solids", edited by M.~L.~Klein and J.~A.~Venables (Academic Press, New York, 1976).
\bibitem{lot00} V. F.~Lotrich and K. Szalewicz, J. Chem. Phys. {\bf 112}, (2000) 112.
\bibitem{whi79} P.~A.~Whitlock, D.~M.~Ceperley, G.~V.~Chester and M.~H.~Kalos, Phys. Rev. B {\bf 19}, (1979) 5598.
\bibitem{coh96} M.~J.~Cohen and J.~N.~Murrell, Chem. Phys. Lett {\bf 260}, (1996) 371.
\bibitem{dut71} N.~C.~Dutta, C.~M.~Dutta and T.~P.~Das, Int. J. Quant. Chem. {\bf 4S}, (1971) 299.
\bibitem{mos82} J.~W.~Moskowitz, K.~E.~Schmidt, M.~A.~Lee and M.~H.~Kalos, J. Chem. Phys. {\bf 77}, (1982) 349.
\bibitem{rey82} P.~J.~Reynolds, D.~M.~Ceperley, B.~J.~Alder and W.~A.~Lester, J. Chem. Phys. {\bf 77}, (1982) 5593.
\bibitem{umr93} C.~J.~Umringar, M.~P.~Nightingale and K.~J.~Runge, J. Chem. Phys. {\bf 99}, (1993) 2865.
\bibitem{cep81} D.~M.~Ceperley, M.~H.~Kalos and J.~L.~Lebowitz, Macro-molecules {\bf 14}, (1981) 1472.
\bibitem{liu74} K.~S.~Liu, M.~H.~Kalos and G.~V.~Chester, Phys. Rev. A {\bf 10}, (1974) 303.
\bibitem{bar99} S. Baroni and S. Moroni, Phys. Rev. Lett. {\bf 82}, (1999) 4745.
\bibitem{att92} P. Attard, Phys. Rev. A {\bf 45}, (1992) 5649.
\bibitem{mur71} R.~D.~Murphy and J.~A.~Barker, Phys. Rev. A {\bf 3}, (1971) 1037.
\bibitem{azi73} R.~A.~Aziz and R.~K.~Pathria, Phys. Rev. A {\bf 7}, (1973) 809.
\bibitem{roa70} P.~R.~Roach, S.~B.~Ketterson and C.~W.~Woo, Phys. Rev. A {\bf 2}, (1970) 543.
\bibitem{sar03} A.~Sarsa and K.~E.~Schmidt, Private Communication.
\bibitem{woo77} A.~D.~B.~Woods and V.~Sears, Phys. Rev. Lett. {\bf 39}, (1977) 415.
\bibitem{mor98} S.~Moroni, D.~E.~Galli, S.~Fantoni and L.~Reatto, Phys. Rev. B {\bf 58}, (1998) 909.

\end{thebibliography}
%\bibliographystyle{prsty}

\newpage

\begin{table}[h] \label{tab:tene}
\centering
\caption{Total energies per atom in units of K obtained at the given densities and potentials. Results in the second column for the potential $V_{2}$, in the third and fourth columns the $V_{2D}$ and $V_{2DJ}$ potentials were considered (see text). In the liquid phase the results were obtained with 64 bodies and in the solid phase with 108 particles. In the last column we show the experimental values.}
\begin{tabular}{|ccccc|}
\hline \hline
$\rho$ $(\rm{nm}^{-3})$ & $E^{(2)}$ & $E^{(2D)}$ &  $E^{(2DJ)}$ & Exp. \\
\hline \hline
& \multicolumn{3}{c}{\rm Liquid} &  \\
19.64 & $-7.121 \pm 0.006$ & $-7.016 \pm 0.006$ & $-7.011 \pm 0.006$ & 
-7.01\footnotemark[2]\\ 
21.86 & $-7.238 \pm 0.009$ & $-7.103 \pm 0.010$ & $-7.097 \pm 0.010$ & 
-7.14\footnotemark[2]\\ 
21.86\footnotemark[1] & $-7.240 \pm 0.007$ & $-7.101 \pm 0.007$ & $-7.095 
\pm 0.007$ & -7.14\footnotemark[2]\\
24.01 & $-7.120 \pm 0.010$ & $-6.949 \pm 0.010$ & $-6.942 \pm 0.010$ & 
-7.00\footnotemark[2]\\ 
26.23 & $-6.541 \pm 0.014$ & $-6.325 \pm 0.014$ & $-6.318 \pm 0.014$ & 
-6.53\footnotemark[2]\\ 
& \multicolumn{3}{c}{\rm Solid} &  \\
29.34 & $-5.907 \pm 0.004$ & $-5.600 \pm 0.003$ & $-5.600 \pm 0.004$ & 
-5.78\footnotemark[3]\\ 
32.88 & $-4.489 \pm 0.006$ & $-4.071 \pm 0.006$ & $-4.076 \pm 0.007$ & \\ 
33.54 & $-4.089 \pm 0.005$ & $-3.648 \pm 0.005$ & $-3.656 \pm 0.005$ & 
-3.94\footnotemark[3]\\ 
35.27 & $-2.831 \pm 0.006$ & $-2.323 \pm 0.006$ & $-2.336 \pm 0.006$ & 
-2.70\footnotemark[3]\\ 
\hline \hline
\end{tabular}
\footnotetext[1]{Result for 108 particles.}
\footnotetext[2]{Reference \onlinecite{azi73}.}
\footnotetext[3]{Reference \onlinecite{woo77}.}
\end{table}

\newpage

\begin{table}[h] \label{tab:taxex}
\centering
\caption{Energies per particle in units of K associated to the triple-dipole term ($E^D$) and the exchange term ($E^J$) at the given densities obtained by reweighting and by extrapolation of perturbative calculation.}
\begin{tabular}{|ccccc|}
\hline \hline
 & \multicolumn{2}{c}{$E^D$} & \multicolumn{2}{c}{$E^J$} \vline \\
$\rho$ $(\rm{nm}^{-3})$ & Rew. & Extr. & Rew. & Extr. \\
\hline \hline
& \multicolumn{3}{c}{\rm Liquid} &  \\
19.64 & $0.105 \pm 0.001$ & $0.1012 \pm 0.0002$ & $0.0044 \pm 0.0003$ & 
$0.0056 \pm 0.0001$ \\
21.86 & $0.135 \pm 0.001$ & $0.1333 \pm 0.0002$ & $0.0056 \pm 0.0004$ & 
$0.0066 \pm 0.0001$ \\
21.86\footnotemark[1] & $0.139 \pm 0.002$ & $0.1351 \pm 0.0006$ & $0.0058 \pm 
0.0004$ & $0.0068 \pm 0.0001$ \\
24.01 & $0.171 \pm 0.001$ & $0.1698 \pm 0.0002$ & $0.0063 \pm 0.0003$ & 
$0.0074 \pm 0.0001$ \\
26.23 & $0.217 \pm 0.001$ & $0.2136 \pm 0.0002$ & $0.0069 \pm 0.0005$ & 
$0.0078 \pm 0.0001$ \\
& \multicolumn{3}{c}{\rm Solid} &  \\
29.34 & $0.307 \pm 0.001$ & $0.3035 \pm 0.0002$ & $0.0000 \pm 0.0003$ & 
$0.0019 \pm 0.0001$ \\
32.88 & $0.418 \pm 0.001$ & $0.4140 \pm 0.0002$ & $-0.0051 \pm 0.0007$ 
& $-0.0028 \pm 0.0002$ \\
33.54 & $0.441 \pm 0.001$ & $0.4379 \pm 0.0002$ & $-0.0072 \pm 0.0008$ 
& $-0.0050 \pm 0.0002$ \\
35.27 & $0.508 \pm 0.001$ & $0.5029 \pm 0.0002$ & $-0.0128 \pm 0.0006$ 
& $-0.0097 \pm 0.0002$ \\
\hline \hline
\end{tabular}
\footnotetext[1]{Result for 108 particles.}
\end{table}

\newpage

\begin{table} [h] \label{tab:lsparam}
\centering
\caption{Fitting parameters of the liquid and solid equations of state for three different potentials. In the first line, for both the liquid and solid phase, the two-body potential $V_{2}$ of Aziz et al. (Ref. \onlinecite{azi95}) was used. Then we present results when the three-body Axilrod-Teller term is included in the interacting potential, $V_{2D}$. In the rows with $V_{2DJ}$ we show results obtained when the full potential, that includes the three-body exchange term in $V_{2D}$, was used. The units of $E_0$, $B$ and $C$ are expressed in K.}
\begin{tabular}{|ccccc|}
\hline\hline
Potential & $\rho_0$ $(\rm{nm}^{-3})$ & $E_0$ & $B$ & $C$ \\
\hline\hline
& \multicolumn{3}{c}{\rm Liquid} &  \\
$V_2$ & 22.133 & -7.240 & 13.549 & 37.025 \\
$V_{2D}$ & 21.845 & -7.103 & 12.143 & 35.705 \\
$V_{2DJ}$ & 21.834 & -7.097 & 12.081 & 35.480 \\
& \multicolumn{3}{c}{\rm Solid} &  \\
$V_2$ & 26.795 & -6.200 & 31.880 & 5.661 \\
$V_{2D}$ & 26.399 & -5.980 & 29.739 & 7.870 \\
$V_{2DJ}$ & 26.045 & -6.028 & 25.233 & 11.844 \\
\hline\hline
\end{tabular}
\end{table}

\newpage

\begin{table}[h] \label{tab:dmf}
\centering
\caption{Melting and freezing densities using three different potentials calculated by the Maxwell double tangent construction method. The $V_2$ potential is the two-body potential of Aziz et al. (Ref. \onlinecite{azi95}). The $V_{2D}$ potential is build using the $V_2$ potential plus the three-body Axilrod-Teller interaction term. Finally the $V_{2DJ}$ potential includes the $V_2$ potential, the three-body Axilrod-Teller and the exchange terms. In the last line we give the experimental values.}
\begin{tabular}{|ccc|}
\hline \hline
Potential & $\rho_{\rm f}$ $(\rm{nm}^{-3})$ & $\rho_{\rm m}$ 
$(\rm{nm}^{-3})$ \\
\hline \hline
$V_2$ & 24.94 & 29.39 \\
$V_{2D}$ & 25.00 & 29.35 \\
$V_{2DJ}$ & 24.99 & 29.28 \\
\rm{Exp.} & 25.8\footnotemark[1] & 28.0\footnotemark[1] \\
\hline \hline
\end{tabular}
\footnotetext[1]{Reference \onlinecite{mor98}.} 
\end{table}

\newpage

\section*{CAPTIONS}

{\bf Figure 1.-} Energy per atom associated to the three-body Axilrod-Teller interaction term for the liquid and solid phases. The crosses stand for the DMC results with reweighting. The circles show extrapolated estimates of perturbative calculations. The statistical errors are smaller than the size of the symbols. The results were obtained using a simulation cell with 64 particles for the liquid phase and 108 for the solid one.

{\bf Figure 2.-} Energy per particle associated to the three-body exchange term for the liquid and solid phases, DMC with reweighting (crosses) and extrapolated perturbative results (circles). The statistical errors of the last calculations are smaller than the size of the symbols. The results were obtained using a simulation cell with 64 particles for the liquid phase and 108 for the solid one.

{\bf Figure 3.-} Analytical equations of state with three different potentials for the solid and liquid phases. The doted line represent the equation of state obtained using the results determined with the two-body potential $V_2$. The solid line represent results using the $V_{2D}$ and $V_{2DJ}$ potentials that includes only the Axilrod-Teller term and this term plus the exchange one, respectively. At the figure scale the two fits are indistinguishable. The points represent results from our calculations.

\newpage

\begin{figure} [h] \label{fig:fat}
\section*{Figure 1}
\begin{center}
\includegraphics[width=15cm,height=15cm]{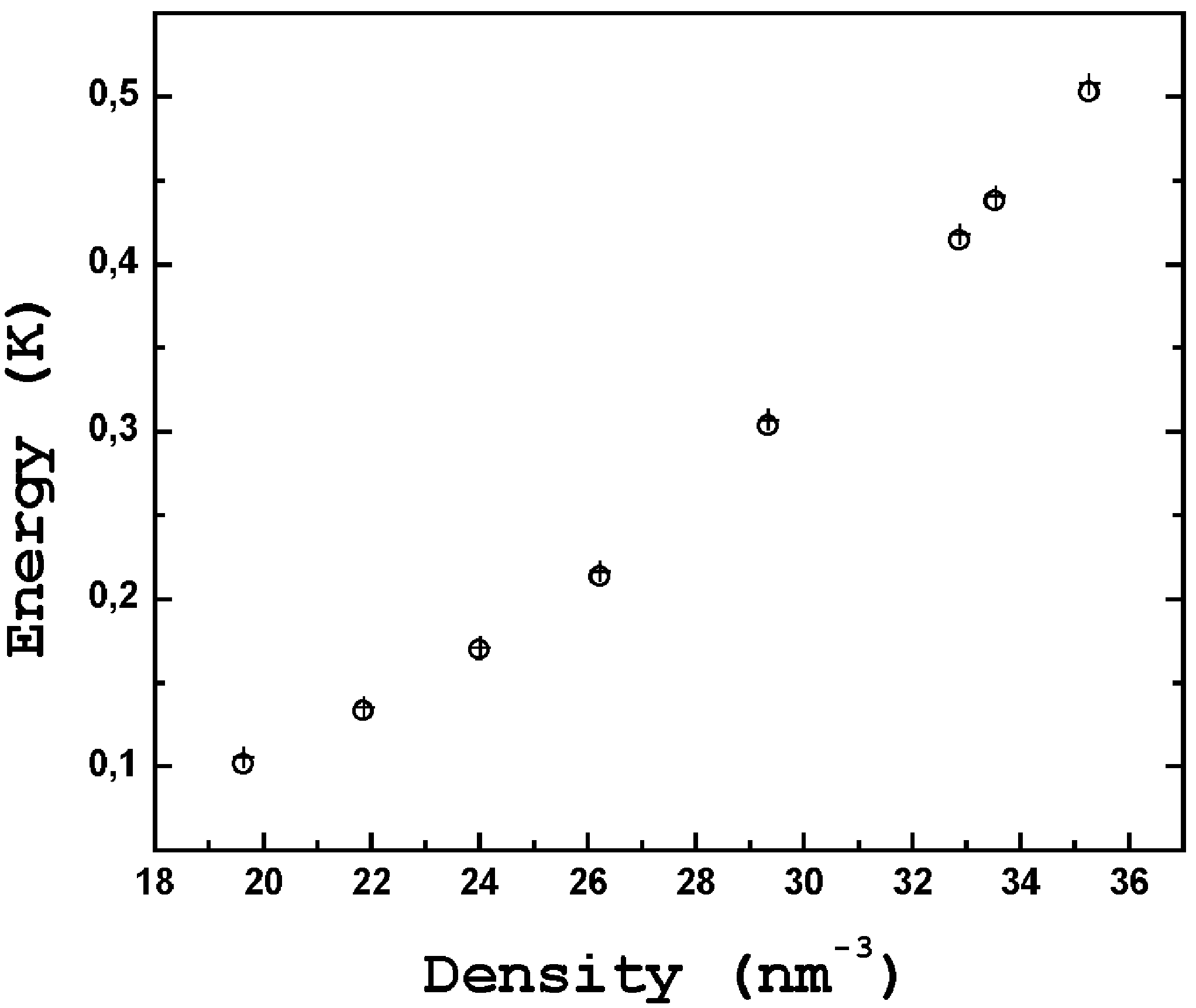}
\end{center}
\end{figure}

\newpage

\begin{figure} [h] \label{fig:fex}
\section*{Figure 2}
\begin{center}
\includegraphics[width=15cm,height=15cm]{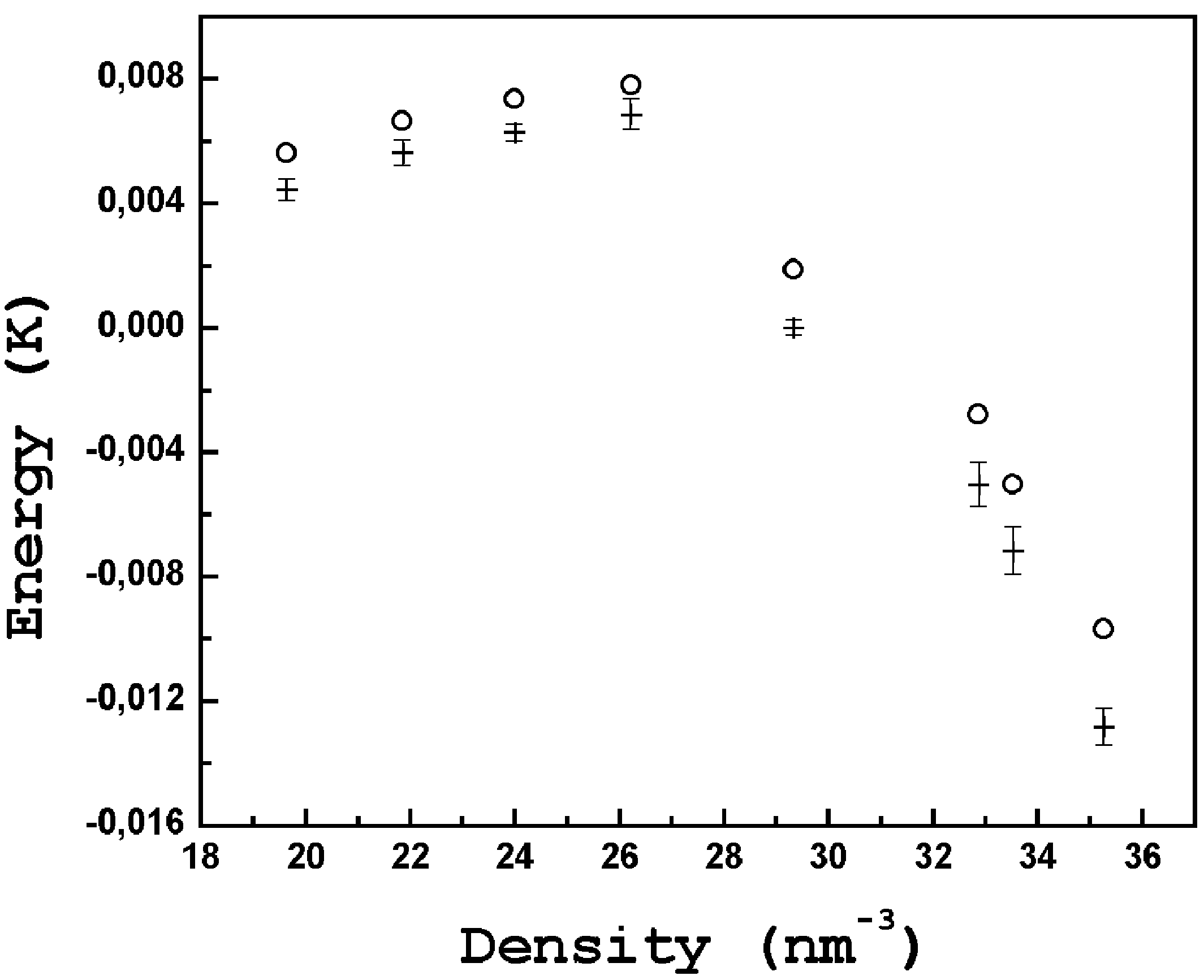}
\end{center}
\end{figure}

\newpage

\begin{figure} [h] \label{fig:eq2dj}
\section*{Figure 3}
\begin{center}
\includegraphics[width=15cm,height=15cm]{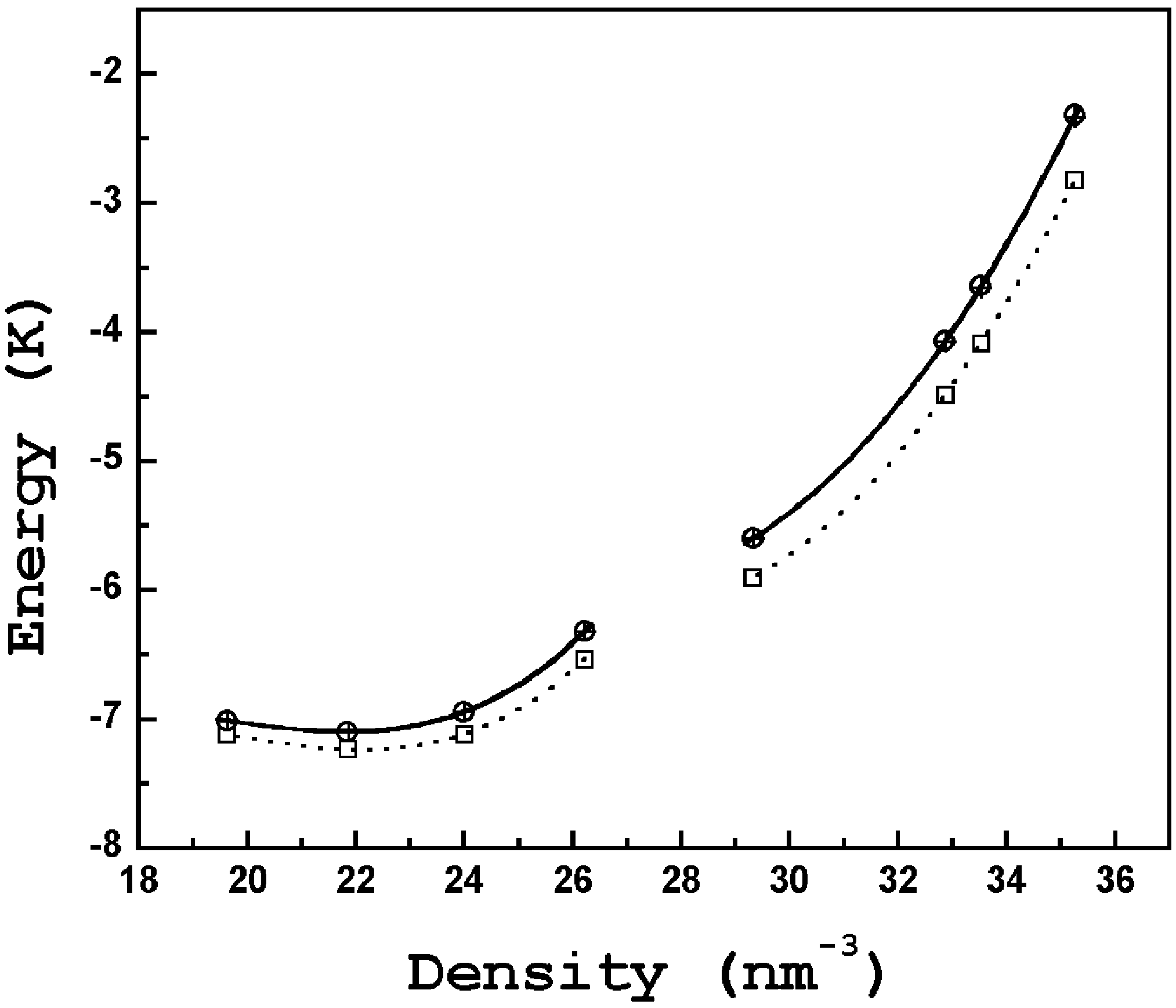}
\end{center}
\end{figure}

\end{document}